# Notes on phonological based drunken detection algorithm

# Oct 2015

Adel Rahimi

Rahimi.adel@gmail.com

In this paper we propose a new algorithm for detecting if a person is under the influence of alcohol. This algorithm is based on number of pauses the speaker makes by judging that if the number of pauses compared to the previous recordings of the same person, which has been recorded beforehand, is higher. If so the algorithm mark the speaker as drunk.

## 1  Past works

There are numerous researches conducted on effects of alcohol, all suggesting fundamental frequency and speech duration are among top factors that change while being intoxicated is the duration of speech i.e. number of pauses.

### a. Number of pauses

Tisljár-Szabó et al (2014), Braun et al (2003) all suggest that the number of pauses will increase while being under influenced. Braun (2003) recorded 33 male subject's speech. He then calculated the total duration of the utterances before and after being intoxicated. Below is the chart from Braun (2003).

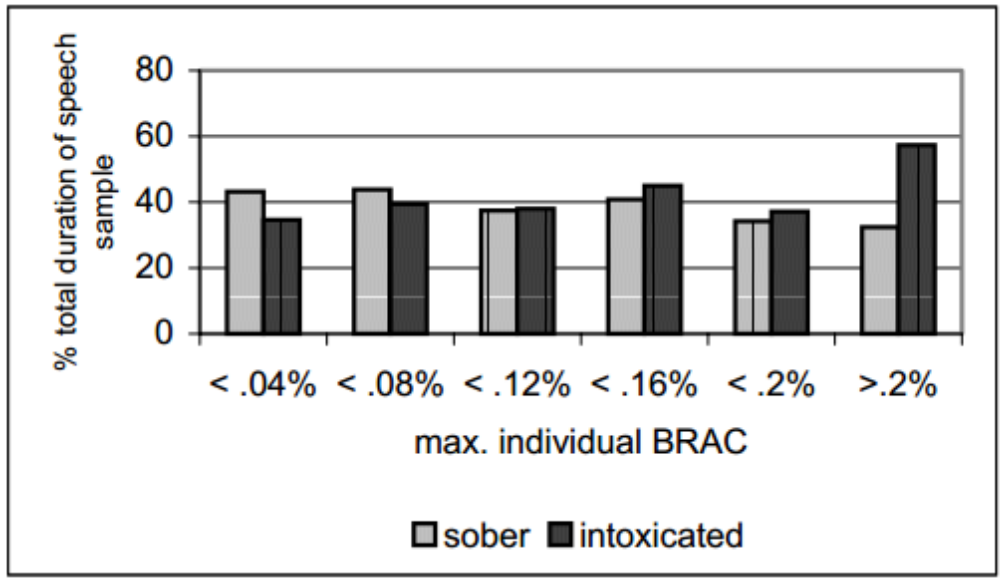

*Figure 1 Percentage of pauses in sober and intoxicated (from Braun 2003)*

This chart shows the correlation of being intoxicated with the number of pauses the speaker makes.

This solely can help on building the algorithm while the program can be trained on several sentences produced by the speaker and the total duration and pauses will be counted and help the algorithm to detect if the talker is intoxicated or not.

We can train our program to cutoff frequencies below 80 Hz and above 300 Hz since our talkers do not sing and therefore we don't need the extra noise on our data. The demo code on cutoff for frequencies on python is below:

```python
import scipy.fftpack as sf
import numpy as np
def maxFrequency(X, F_sample, Low_cutoff=80, High_cutoff= 300):
    """ Searching presence of frequencies on a real signal using FFT
    Inputs
    =======
    X: 1-D numpy array, the real time domain audio signal (single channel time series)
    Low_cutoff: float, frequency components below this frequency will not pass the filter (physical frequency in unit of Hz)
    High_cutoff: float, frequency components above this frequency will not pass the filter (physical frequency in unit of Hz)
```

```
        F_sample: float, the sampling frequency of the signal (physical
frequency in unit of Hz)
        """
```

   b. Fundamental Frequency

When talking about fundamental frequency most researches suggest that the mean fundamental frequency does not change while being intoxicated; however this might be tricky since fundamental frequency is influenced by numerous factors.

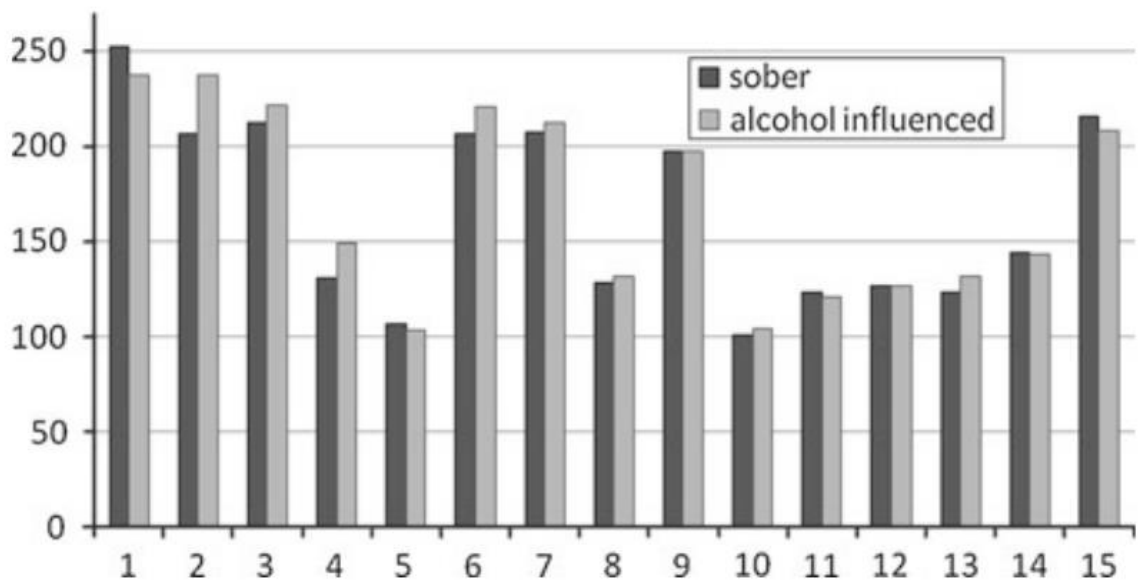

*Figure 2 Mean fundamental frequency by individual speaker in sober and alcohol conditions (from Tisljár-Szabó 2014)*

As Chin suggests imbibing alcohol does effect the number of pauses in speech and also it increases the variability of fundamental frequency however in the paper Chin suggests that the fundamental frequency does not change in the speaker but looking at the charts only speakers number 2 and 3 show this "anomaly". There was no background on speakers. I would suggest there are more factors concerning effects of alcohol on speech and especially fundamental frequency. Factors such as: age, sex, previous medical problems etc. we will get into that in the next section.

## 2 the algorithm proposed

Application of this algorithm based on these three factors the number of pauses the speaker makes

The algorithm can first analyze the frequency of the background noise input between 80 to 300 Hz then by analyzing the intensity of the noise we can get a fingerprint of the fundamental noise, then any sound increasing the intensity in any frequency ranging from 80 to 300 Hz will be detected as speaking and therefore pauses will be detected.

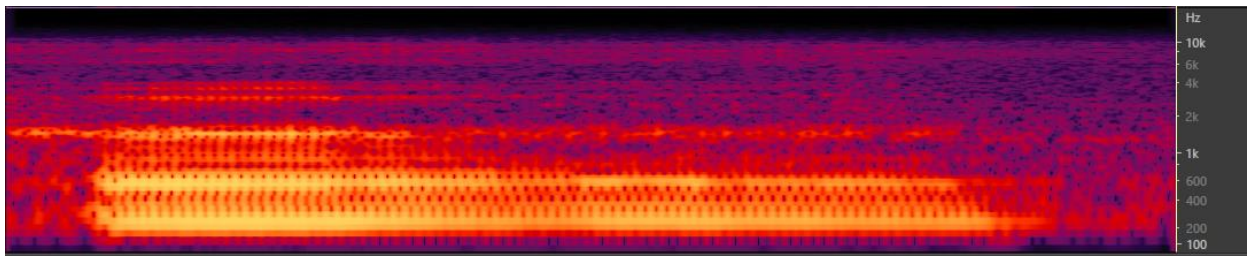

*Figure 3 the spectrogram display showing pause before and after the vowel /e/*

The sample code on python will look like this:

```python
import scipy.fftpack as sf
import numpy as np
def maxFrequency(X, F_sample, Low_cutoff=80, High_cutoff= 300):
    """ Searching presence of frequencies on a real signal using FFT
    Inputs
    =======
    X: 1-D numpy array, the real time domain audio signal (single channel time series)
```

```python
        Low_cutoff: float, frequency components below this frequency will not
pass the filter (physical frequency in unit of Hz)
        High_cutoff: float, frequency components above this frequency will not
pass the filter (physical frequency in unit of Hz)
        F_sample: float, the sampling frequency of the signal (physical
frequency in unit of Hz)
        """
        M = X.size # let M be the length of the time series
        Spectrum = sf.rfft(X, n=M)
        [Low_cutoff, High_cutoff, F_sample] = map(float, [Low_cutoff,
High_cutoff, F_sample])
        
        #Convert cutoff frequencies into points on spectrum
        [Low_point, High_point] = map(lambda F: F/F_sample * M, [Low_cutoff,
High_cutoff])

        maximumFrequency = np.where(Spectrum == np.max(Spectrum[Low_point :
High_point])) # Calculating which frequency has max power.

        return maximumFrequency

voiceVector = []
for window in fullAudio: # Run a window of appropriate length across the audio
file
    voiceVector.append (maxFrequency( window, samplingRate))
```

# 3  Conclusions

This method can help on using a small amount of resources and in the meantime gives a superb output. It can be used in DUI prevention systems in vehicles, security systems in offices, and mobile phones.